\documentclass[11pt,a4paper]{article}
\pdfoutput=1
\usepackage{jinstpub}
\usepackage{multirow}
\usepackage{enumitem}
\numberwithin{table}{section}

\title{Muon Counting using Silicon Photomultipliers in the AMIGA detector of the Pierre Auger Observatory}

\author{The Pierre Auger collaboration}
\emailAdd{auger\_spokespersons@fnal.gov}

\abstract{AMIGA (Auger Muons and Infill for the Ground Array) is an upgrade of the Pierre Auger Observatory designed to extend its energy range of detection and to directly measure the muon content of the cosmic ray primary particle showers. The array will be formed by an infill of surface water-Cherenkov detectors associated with buried scintillation counters employed for muon counting. Each counter is composed of three scintillation modules, with a 10\,m$^2$ detection area per module. In this paper, a new generation of detectors, replacing the current multi-pixel photomultiplier tube (PMT) with silicon photo sensors (aka. SiPMs), is proposed. The selection of the new device and its front-end electronics is explained. A method to calibrate the counting system that ensures the performance of the detector is detailed. This method has the advantage of being able to be carried out in a remote place such as the one where the detectors are deployed. High efficiency results, i.e. 98\,\% efficiency for the highest tested overvoltage, combined with a low probability of accidental counting ($\sim$2\,\%), show a promising performance for this new system.  
}

\keywords{Performance of High Energy Physics Detectors; Photon detectors for UV, visible and IR photons (solid-state) (PIN diodes, APDs, Si-PMTs, G-APDs, CCDs, EBCCDs, EMCCDs etc); Front-end electronics for detector readout; Pattern recognition, cluster finding, calibration and fitting methods}

\begin{document}
\maketitle
\flushbottom
\newpage
\section{Introduction}
\label{sec:Intro}

The Pierre Auger Obsevatory~\cite{pao_nim} is located in the province of Mendoza, Argentina and has an area of 3000\,km$^2$. It was designed to detect ultra-high energy cosmic ray showers with a hybrid detection technique. It has 1660 surface water-Cherenkov detector stations (SDs)~\cite{Auger-SD08} arranged in a triangular grid with a distance of $\sim$1.5\,km between stations, and 27 fluorescence detector (FD) telescopes~\cite{Auger-FD10} at four sites on the periphery of the array pointing towards the atmosphere and the center of the array. The Auger Observatory is currently being upgraded, and AMIGA~\cite{Etchegoyen07, Platino09, Buchholz09} (Auger Muons and Infill for the Ground Array) is one of its principal enhancements. Two of the main objectives of AMIGA are the measurement of composition-sensitive observables of extensive air showers and the study of features of hadronic interactions. Important results on cosmic ray physics by means of muon detection techniques have been previously obtained by several experiments like KASCADE~\cite{KASCADE} and KASCADE-Grande~\cite{KASCADE-Grande}.

AMIGA consists of 61 detector pairs, each one composed of a SD station and a 30\,m$^2$ muon counter, deployed on a 750\,m triangular grid in an infilled area of 23.5\,km$^2$. At the Observatory site, the associated SD triggers the muon counter (MC) when a candidate cosmic ray shower is measured. Each muon counter is buried 2.3\,m underground to shield the electromagnetic component of cosmic ray showers (vertical shielding of 540\,g/cm$^2$) and it is composed of three scintillation modules. Every module comprises 64 scintillation bars, each of dimensions 400\,cm x 4\,cm x 1\,cm, with a 1.2\,mm diameter wavelength-shifting (WLS) optical fiber (BCF-99-29AMC) glued to a lengthwise groove on each bar. The light produced in the bars is absorbed by the WLS fiber. The excited molecules of the fiber decay while emitting photons, some of which are propagated along the WLS fibers towards a channel of a multi-pixel photon detector. The aim of these modules is to efficiently count the number of muons that impinge on the 10\,m$^2$ area of scintillation. The mechanical detector design is fixed and detailed in~\cite{Suarez01}. An engineering array of seven muon counters with multi-pixel photomultiplier tubes (PMTs) is already installed and acquiring data at the Observatory site.  

For the production of AMIGA the current PMTs are going to be replaced with silicon photo sensors (aka. SiPMs). 
The main motivations for this upgrade are the advantages of these devices compared to current PMTs: their lower cost per channel, longer life-time, better sturdiness, higher photon detection efficiency at the optical fiber emission wavelength, and no optical cross-talk between channels. The main disadvantages of SiPMs are its higher noise rate and temperature dependence. In this paper is explained how, with the proposed electronics and calibration method, these disadvantages are addressed. 

The present paper is organized as follows: a general description of the SiPM behavior is detailed in section~\ref{sec:The Silicon Photomultiplier}. Then the SiPM and the front-end electronics selection is explained in section~\ref{sec:Proposed Readout for AMIGA Muon Counters}. The proposed calibration of the counting system is described in section~\ref{sec:Calibration Method of the Counting System}. Finally, the efficiency measurements are shown in section~\ref{sec:Efficiency Measurements}. 

\newpage

\section{The Silicon Photomultiplier}
\label{sec:The Silicon Photomultiplier}

\begin{figure}[h!t]
	\centering
	\includegraphics[width=0.4\textwidth]{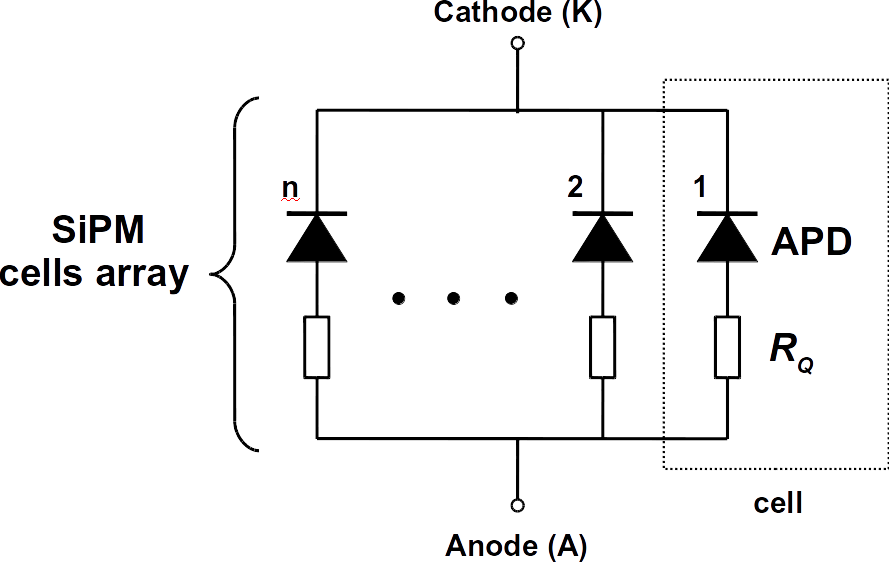}
	\caption{Schematic view of the internal structure of a SiPM made up of an array of cells, all connected in parallel. Each cell is composed of an APD working in Geiger mode and a quenching resistor ($R_{Q}$) in series.}
	\label{fig:array}
\end{figure}

A SiPM~\cite{HamamtsuSiPM} is a solid-state device capable of detecting individual photons. It is composed of an array of cells, all connected in parallel. Each cell has an avalanche photo-diode (APD) working in Geiger mode and a quenching resistor ($R_{Q}$) in series (see figure~\ref{fig:array}). 

The APD starts working in Geiger mode when the reverse voltage ($V_{bias}$) applied to it exceeds a specific voltage value called the breakdown voltage ($V_{BR}$). In this mode the injection of a single charge carrier (e.g. due to an impinging photon) causes a self-sustained avalanche. The current that flows through the APD depends on the voltage value over the breakdown which is called overvoltage ($\Delta V = V_{bias} - V_{BR}$). The flow of this current through $R_{Q}$ produces the decrease of the reverse voltage ($V_{APD}$) applied to the APD. When $V_{APD}$ is below $V_{BR}$ the avalanche is extinguished. This last sequence describes the ``firing'' of a cell. 
From now on, the signal produced by this firing process will be called \textit{single photon equivalent} (SPE). If multiple cells are fired simultaneously the resulting output signal will be a superposition of SPEs. The amplitude of this signal will be directly proportional to the number of fired cells. 

As was previously mentioned, SiPMs are employed to detect photons. An undesirable effect is the accidental counting of those photons due to noise in the cells. In the next subsection, the sources of noise are defined. 

\subsection{SiPM Noise}
\label{subsec:SiPM Noise}

Noise in SiPMs~\cite{Characterisation SiPMs} is defined as the firing of a cell that was not produced by a photon impinging the device. There are three noise sources which can be separated in two types, depending on the correlation or not with the firing of a cell. 
	
\subsubsection{Uncorrelated Noise: Dark Noise}

Dark noise occurs randomly due to thermally-generated charge carriers (electron-hole pairs) either in the depletion region or in the avalanche region~\cite{Depletion region}. The amplitude and shape of these pulses are the same as the ones produced by the absorption of a photon. Dark noise is sensitive to temperature, and also depends on the array size, overvoltage magnitude, and semiconductor material quality.
	
\subsubsection{Correlated Noise: Afterpulsing and Cross-Talk}

Afterpulsing is a secondary avalanche produced after the firing of a cell, due to the release of trapped charges. The release of these trapped charges occurs after a characteristic time that depends on the type of the trapping centers and its occurrence probability decreases exponentially with time. It is noise correlated to the firing of a cell and it is produced in the same cell.

When a primary avalanche in a cell produces photons with energy greater than the band gap energy, there is a probability that a nearby cell absorbs the photon, producing its firing. As a first order approximation, the secondary avalanche is synchronized in time with the main primary avalanche to produce a resulting signal of a channel with an increased amount of SPEs stacked. This effect is called cross-talk.

\section{Proposed Readout for AMIGA Muon Counters}
\label{sec:Proposed Readout for AMIGA Muon Counters}

As mentioned in the Introduction, for the production of AMIGA, the current PMTs are going to be replaced with SiPMs. The mechanical module design is already fixed. 
One of the characteristics that constrains the proposed readout for AMIGA muon counters is its segmentation. The detector is segmented, in space as well as in time, to prevent undercounting due to simultaneous muon arrivals:

\begin{itemize}
\item \textit{Segmentation in space.} Based on simulations~\cite{Supanitsky08}, each of the three scintillator modules of a muon counter has been segmented into 64 segments.  
\item \textit{Segmentation in time.} This segmentation is limited by the time distribution of the photons produced by the impinging particle. This time distribution is defined by the convolution of the probability distributions characterized by the decay time of the scintillator, the decay time of the optical fiber, and to a lesser extent the propagation mode in the optical fiber. The maximum time width of a light signal for AMIGA MCs is between 25 to 35\,ns~\cite{ICRCWundheiler}. 
\end{itemize}

The proposed electronics of the module must facilitate the identification of pulses above a given threshold to allow muon counting, without knowing in detail the signal structure and peak intensity.
The proposed readout must not degrade the detector segmentation. 

A SiPM model and new electronics for the readout of AMIGA MCs are proposed in the next two subsections, based on the experience of the current version of the AMIGA electronics~\cite{OWelectronica, ICRCAlmela}. 

\subsection{SiPM Selection}

Two main features which improve the signal-to-noise ratio were taken into account in order to select the specific device: high photo-detection efficiency (PDE) and low noise. The PDE of the selected devices is around 35\,\% for the peak emission wavelength of the fiber optic (485\,nm). Low noise is obtained by combining low dark rate with reduced cross-talk and low afterpulsing probability. 

\begin{figure}[h!t]
	\centering
	\includegraphics[width=\textwidth]{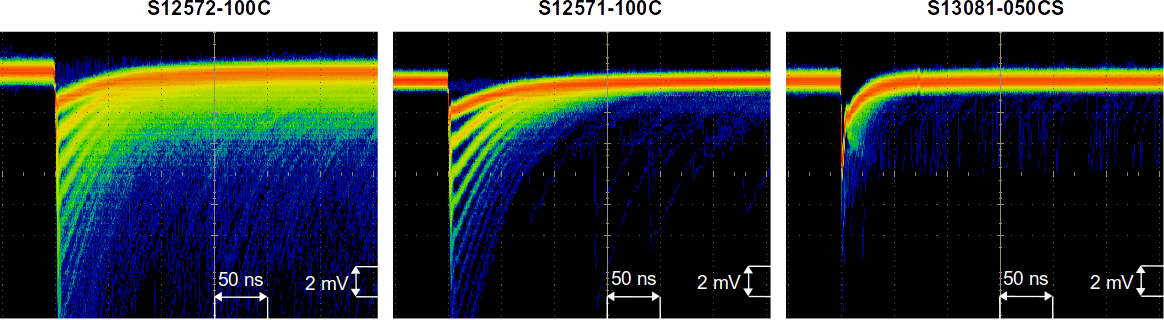}
	\caption{Overlap of 5000 dark rate traces (signal amplitude as a function of time) of each SiPM model. All measurements were done with the same amplifier at 25\,$^\circ$C. The $\Delta V$ of each SiPM was set to the value recommended by Hamamatsu.}
	\label{fig:SiPM comparacion}
\end{figure}

Three devices manufactured by Hamamatsu (S12572-100C, S12571-100C, S13081-050CS) were tested in the laboratory to evaluate their performance. In Figure~\ref{fig:SiPM comparacion} an overlap of 5000 dark rate traces of each SiPM model is shown. Pulses of more than one SPE, stacked due to cross-talk, can be observed synchronized with the trigger time. Afterpulsing pulses can also be observed after the trigger time. 

The three SiPMs exemplified are some of the latest devices developed by Hamamatsu up to 2015. The second SiPM (S12571-100C) in figure~\ref{fig:SiPM comparacion} shows a reduction in the afterpulsing probability compared to the first one (S12572-100C). The third model (S13081-050CS) not only shows a reduction in the afterpulsing, but also a significantly lower value of cross-talk. The main characteristics of these SiPMs, obtained from the Hamamtsu datasheets, are summarized in table~\ref{tab:SiPM comparative}.

The criteria selected for counting muons is to count pulses above a given threshold. In this context, cross-talk probability becomes a relevant parameter. This correlated noise makes possible that a pulse triggered by dark noise or afterpulsing could have an amplitude above the selected threshold. Therefore, cross-talk should be as small as possible to reduce the accidental counting probability. The selected device for this work was the third model (S13081-050CS) mainly due to its low cross-talk probability and also its low dark noise and afterpulsing.

\begin{table}[h!]
\centering
\caption{Main characteristics, obtained from the Hamamatsu datasheets, of the three tested SiPMs: S12571-100C~\cite{S12571-100C}, S12572-100C~\cite{S12572-100C} and S13081-050CS.}
\vspace{0.3cm}
\begin{tabular}{|c|c|c|c|c|c|}
\hline
\multicolumn{2}{|c|}{\multirow{2}{*}{\textbf{Parameter}}} & \multicolumn{3}{c|}{\textbf{SiPM Model}}                                               & \multirow{2}{*}{\textbf{Unit}} \\ \cline{3-5}
\multicolumn{2}{|c|}{}                                    & S12572-100C                & S12571-100C                & S13081-050CS               &                                \\ \hline
\multicolumn{2}{|c|}{Cell Pitch}                         & 100                        & 100                        & 50                         & $\mu$m                         \\ \hline
\multicolumn{2}{|c|}{Effective Photosensitive Area}       & 3 x 3                      & 1 x 1                      & 1.3 x 1.3                  & mm                             \\ \hline
\multicolumn{2}{|c|}{Geometrical Fill Factor}             & 78.5                       & 78                         & 61                         & \%                             \\ \hline
\multicolumn{2}{|c|}{Photon Detection Efficiency}         & 35                         & 35                         & 35                         & \%                             \\ \hline
\multicolumn{2}{|c|}{Number of Cells}                    & 900                        & 100                        & 667                        & -                              \\ \hline
\multirow{2}{*}{Dark Count}             & Typ.            & 1000                       & 100                        & 90                         & kcps                           \\ \cline{2-6} 
                                        & Max.            & 2000                       & 200                        & 360                        & kcps                           \\ \hline
\multicolumn{2}{|c|}{Gain M (at operation voltage)}  & 2.8 x 10$^6$ 				 & 2.8 x 10$^6$ 				 & 1.5 x 10$^6$ 
& -                              \\ \hline
\multicolumn{2}{|c|}{Gain Temperature Coefficient}  & 1.2 x 10$^5$ 				 & 1.2 x 10$^5$ 				 & 2.7 x 10$^4$ 
& $^{\circ}C^{-1}$                              \\ \hline
\multicolumn{2}{|c|}{Breakdown Voltage}  & 65 $\pm$ 10 				 & 65 $\pm$ 10 				 & 53 $\pm$ 5 
& V                              \\ \hline
\multicolumn{2}{|c|}{Cross-Talk Probability}               & 35                         & 35                         & 1                          & \%                             \\ \hline
\multicolumn{2}{|c|}{Temperature Coefficient}               & 60                         & 60                         & 54                          & mV$/^{\circ}C$                             \\ \hline
\end{tabular}
\label{tab:SiPM comparative}
\end{table}

\subsection{Proposed Electronics}

The proposed electronics is based on the existing AMIGA system. The following main specifications were considered for the design:

\begin{enumerate}
\item The new electronics design must be compatible with the existing mechanical design of the detector. There is no access to the optical fibers, only to the optical connector. Fibers of 1.2\,mm diameter are glued to the optical connector in a squared arrangement of 8$\times$8 and the separation distance between two neighboring fibers centers is 2.3\,mm. 
\item The electronics must be able to identify light pulses with a maximum time width between 25 to 35\,ns. This upper limit is obtained from the light pulses time width distribution. This distribution is the result of the convolution of the probability distributions characterized by the decay time of the scintillator, the decay time of the optical fiber, and to a lesser extent the propagation mode in the optical fiber.
\item Low power consumption (stand-alone power system).
\item The output of each SiPM channel must be a digital 0-1 signal to allow muon counting. The digital output width must be similar to the characteristic time width of the light pulses produced in the detector. Wider digital signals are not desired since the detection of consecutive muons in the same bar could be deteriorated (pile-up effect). 
\item Temperature compensation (SiPM breakdown voltage depends on temperature). 
\end{enumerate}

The \textit{Application Specific Integrated Circuit} (ASIC) \textit{Cherenkov Imaging Telescope Integrated Read Out Chip} (CITIROC)~\cite{OmegaCITIROC} was proposed for the front-end readout of AMIGA. 
The CITIROC ASIC is a 32 channel front-end specially designed for the readout of SiPMs. An adjustment of the SiPM biasing is possible using a channel-by-channel 8-bit digital-to-analog converter (DAC) connected to the ASIC inputs. 
Each channel has a pre-amplifier stage that can be selected by software between a low or high gain pre-amplifier. Also each corresponding gain value is programmable. Then the signal passes through a 15\,ns peaking time fast shaper followed by a discriminator. 
The discriminator threshold is set coarsely by a 10-bit DAC (common for the 32 channels) and then set finely channel by channel by individual 4-bit DACs. 
The 15\,ns peaking time fast shaper enables a digital output width of the discriminator similar to the characteristic time width of the light pulses produced by the impinging particles in the detector.  
The fast shaper produces an undesirable effect. It has a negative overshot which might influence the detection efficiency for the next close in time muon. The capability to detect two consecutive muons is not within the scope of this work. This effect and others like larger signal time width, due to the WLS fiber decay time, and the recovery time of cells in the SiPM, must be studied further in the future. 

For the biasing and temperature compensation of each SiPM, the Hamamatsu C11204-01 power supply~\cite{C11204-01} was selected due to the recommendation of Hamamatsu. This power supply has one output with low ripple noise (0.1\,mVp-p typ.), good temperature stability ($\pm$10\,ppm/$^{\circ}$C typ.) and a high output voltage resolution (1.8\,mV).

\section{Calibration Method of the Counting System}
\label{sec:Calibration Method of the Counting System}

The calibration method of the counting system for AMIGA is split into two steps. The first step consists in calibrating the optical sensors of each individual channel (see subsection~\ref{subsec:CalibSiPM}) and the second step consists in calibrating the detector (see subsection~\ref{subsec:CalibDetector}).

\subsection{SiPM Calibration}
\label{subsec:CalibSiPM}

The goal of this calibration is to set the operation point of the SiPMs. First, the breakdown voltage of each individual channel must be obtained. Then, all the SiPMs must be biased to its corresponding breakdown voltage with an added pre-determined overvoltage. This overvoltage can be changed to optimize the efficiency. 
In this section a method for obtaining the breakdown voltage and biasing of the SiPMs with the proposed AMIGA electronics is explained.

\subsubsection{SiPM Calibration Setup}
\label{subsec:Calib_Setup_SiPM}

\begin{figure}[h!t]
	\centering
	\includegraphics[width=\textwidth]{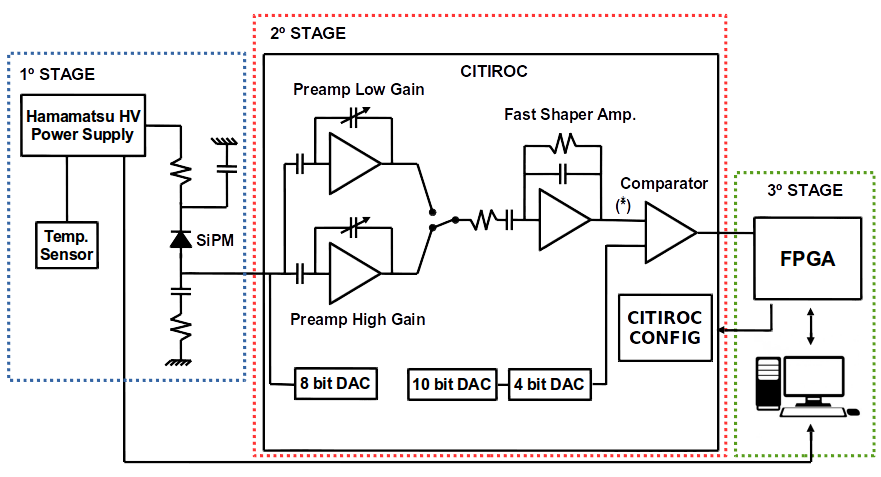}
	\caption{SiPM Calibration Setup. The three stages are distinguished with dotted lines. The output of the fast shaper amplifier ($\ast$) has a DC offset component.}
	\label{fig:sipm_calibration_setup}
\end{figure}

The setup for the SiPMs calibration is divided into three stages (see figure~\ref{fig:sipm_calibration_setup}). 
The first stage is composed of the SiPM, the high voltage power supply and the temperature sensor. Since the SiPM breakdown voltage varies significantly with temperature, the high voltage power supply has a built-in high precision temperature compensation system that constantly corrects the SiPM operation point. This function tries to keep the gain value fixed independently of temperature variations. The compensation of the high voltage ($HV$) output of the power supply is determined by the following equation given by Hamamatsu: 
\begin{equation}
HV = K_{2} * (T - T_{ref})^2 + K_{1} * (T-T_{ref}) + V_{ref}
\label{eq:temperature_compensation}
\end{equation}
In this formula, the $K_{2}$ and $K_{1}$ are respectively the quadratic and linear coefficients for the temperature compensation, $V_{ref}$ is the reference voltage and $T_{ref}$ is the reference temperature. The temperature  compensation for the SiPMs under test is linear and the quadratic term is not needed. The $K_{2}$ was set to 0\,mV/$^{\circ}$C$^2$, $K_{1}$ to 54\,mV/$^{\circ}$C (see table~\ref{tab:SiPM comparative}) and $T_{ref}$ to 25\,$^{\circ}$C. The resulting formula for the calibration is: 
\begin{equation}
HV = 54\,mV/^{\circ}C * (T-25\,^{\circ}C) + V_{ref} 
\label{eq:final_temperature_compensation}
\end{equation}

The second stage consists of the CITIROC. This stage amplifies and then discriminates SiPM pulses. The chip was programmed to use the high gain pre-amplifier, with its maximum gain value of 150, to improve the separation between SPEs in the SPE spectrum. The 8-bit DAC was set to a fixed value (e.g. 250 dac-units). 
The 10-bit DAC was used to set the comparator threshold for all the channels and the 4-bit DAC of each channel was fixed to its minimum value.

The third stage is composed of an Altera Cyclone IV FPGA. The FPGA was programmed to measure the rate of the CITIROC digital pulses output. 

\subsubsection{Single Photon Equivalent Peak Measurement}
\label{subsubsec:Single Photon Equivalent Peak Measurement}

\begin{figure}[h!t]
	\centering
	\includegraphics[width=\textwidth]{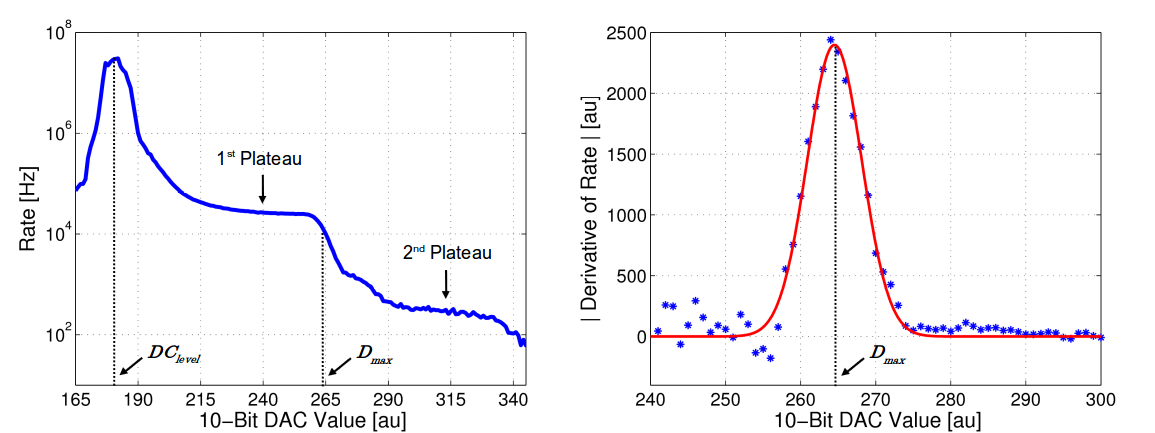}
 	\caption{On the left, the measurement of the rate of the SiPM pulses as a function of the DAC value. $DC_{level}$ corresponds to the fast shaper DC offset component. See subsection~\protect\ref{subsec:Calib_Setup_SiPM} for details on the setup. On the right, the absolute value of the derivative of the rate evidences the distribution of the SPE peaks. The mean value of the SPE peak ($D_{max}$) is obtained with a Gaussian fit of the points (red curve).}
	\label{fig:rate_completo}
\end{figure}

A dedicated software was developed to automatically measure the rate of the digital pulses at different discrimination levels. In the measurement of the SiPM noise rate as a function of the 10-bit DAC values (see figure~\ref{fig:rate_completo}, left) there is a clear transition from the first to the second plateau. This transition is required for the calibration and it represents the threshold of the comparator passing through the first SPE peak.

The absolute value of the derivative of this curve (see figure~\ref{fig:rate_completo}, right) represents the distribution of the SPE peak values. The mean value of the SPE peak ($D_{max}$) is correlated to the maximum absolute value of the Gaussian distribution coming from the derivative of the rate curve. This value has an offset ($DC_{level}$) because the signal in the fast shaper has a DC offset component (see figure~\ref{fig:sipm_calibration_setup}). The value of this DC offset is the DAC value where the rate is maximum (see figure~\ref{fig:rate_completo}, left). 
To obtain the real value of the SPE peak ($SPE_{peak}$) it is necessary to subtract this offset. 
\begin{equation}
SPE_{peak} = D_{max} - DC_{level}
\label{eq:SPE_peak}
\end{equation}

As was mentioned before, the employment of SiPMs combined with the method explained in this subsection, allow the $SPE_{peak}$ estimation which is used to calculate the breakdown voltage of the device, as it will be explained in the next subsection.  

\subsubsection{Breakdown Voltage Measurements}

There are several methods to estimate the breakdown voltage of a SiPM~\cite{breakdown sipms}.
Due to the constraints of the proposed electronics, the method to estimate each channel breakdown voltage is the one described in this subsection. 

The total charge produced in an avalanche can be calculated using equation~\ref{eq:charge} and the gain ($M$) of this process is defined by equation~\ref{eq:gain}. 
	
\begin{equation}
Q = C_j \Delta V
\label{eq:charge}
\end{equation}
\begin{equation}
M = Q/e \hspace{0.5cm} \textrm{where}\ e\ \textrm{is the electron charge.}
\label{eq:gain}
\end{equation}

\begin{figure}[h!t]
	\centering
	\includegraphics[width=\textwidth]{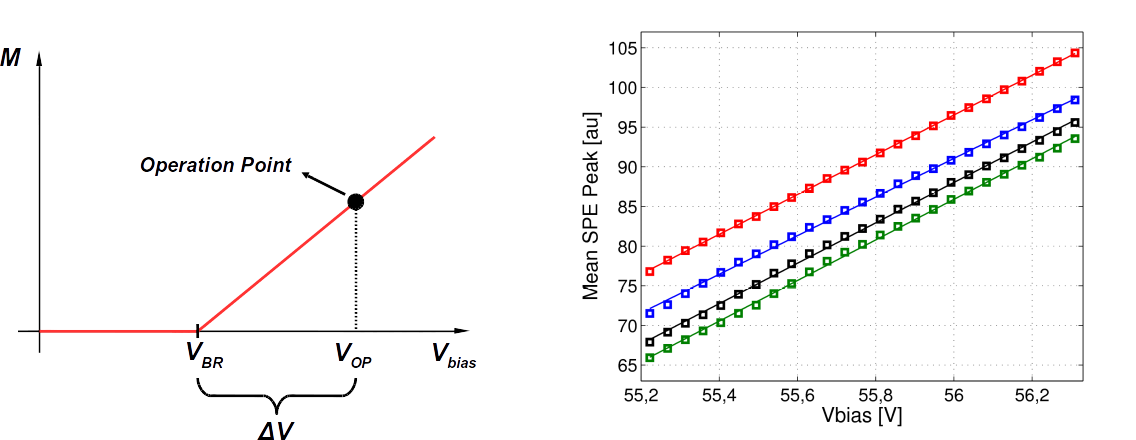}
	\caption{Equations~\protect\ref{eq:charge} and~\protect\ref{eq:gain} are presented graphically in the plot on the left. The mean $SPE_{peak}$ as a function of the $V_{bias}$ for four different SiPMs is plotted on the right. The mean $SPE_{peak}$ is proportional to $M$.}
	\label{fig:breakdown}
\end{figure}

It can be inferred from these equations that the gain depends linearly on the overvoltage. Therefore, for the special case when the gain is zero, $V_{bias} = V_{BR}$. 
The functional dependencies described by these equations are shown in figure~\ref{fig:breakdown}, left. This figure shows that if the gain ($M$) is measured over the $V_{bias}$, the breakdown voltage can be obtained as the value where the curve intercepts the X axis ($V_{bias} = V_{BR}$). It is also known that the $SPE_{peak}$ is directly proportional to the gain (Mean $SPE_{peak}$ $\varpropto$ M). By using this information and following the procedure explained in the previous subsection, a plot of $SPE_{peak}$ for different $V_{bias}$ values, can be obtained. 
An example for four different SiPMs is shown in figure~\ref{fig:breakdown}, right. From that plot, the $V_{BR}$ can be estimated as the point where the linear fit of the curve intercepts the X axis ($V_{bias} = V_{BR}$).

To automatize the breakdown voltage estimation, the 8-bit DAC of the CITIROC (see figure~\ref{fig:sipm_calibration_setup}) was fixed to 250 dac-units for all the channels and the HV value was modified in regular steps ($V_{bias}$ changes following the power supply HV). With this method, each SiPM characteristic breakdown voltage was calculated at the same time, with a single power supply. This is required for the AMIGA design since the 64 channels of each module are connected to the same power supply. In the following subsection a possible equalization of SiPMs overvoltage is explained following this constraint. 

\subsubsection{Equalization Between Channels}
\label{subsec:Equalization Between Channels}

The equalization consists in applying the same $\Delta V$ to all the 64 channels of one module. This equalization does not ensure the same gain or rate at a given threshold between channels. To set the operation voltage for each channel, considering that each SiPM has a different breakdown voltage ($V_{BR_i}$), the following procedure is carried out:
 
\begin{enumerate}
	\item Set the $HV$ voltage of the power supply to the largest $V_{BR_i}$ ($V_{BR_{max}}$) of the 64 SiPMs with the desired $\Delta V$ added. $HV = V_{BR_{max}} + \Delta V$.
	\item Set the 8-bit DAC of each CITIROC channel following: $V_{8bitDAC} = V_{BR_{max}} - V_{BR_i}$.
\end{enumerate}

\subsection{Detector Calibration}
\label{subsec:CalibDetector}

Once the SiPM is calibrated, the next step consists of determining the discrimination level and the counting strategy (detector calibration), ensuring an adequate performance of the counting system.

\subsubsection{Detector Calibration Setup}
\label{subsec:Detector Calibration Setup}

\begin{figure}[h!t]
	\centering
	\includegraphics[width=\textwidth]{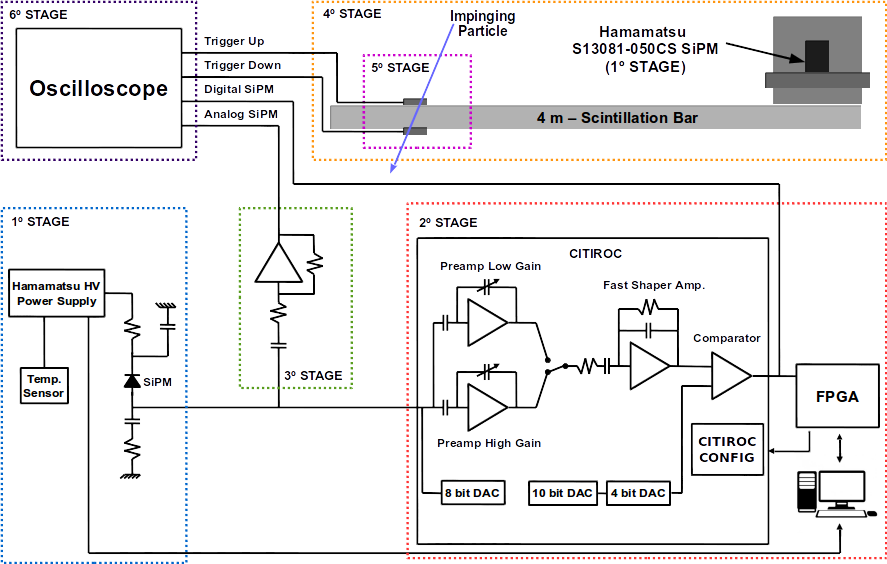}
	\caption{The six stages of the setup needed for the detector calibration. The 4\,m plastic scintillation bar has a threaded 5\,m wavelength-shifting optical fiber. The difference in length is due to the extra fiber needed to reach the optical connector.}
	\label{fig:Detector_Calibration_Setup}
\end{figure}

The setup for the detector calibration is divided into six stages (see figure~\ref{fig:Detector_Calibration_Setup}). The first and second stages were described in subsection~\ref{subsec:Calib_Setup_SiPM}.
For this calibration, the CITIROC was programmed to use the high gain pre-amplifier with its minimum gain value of ten to reduce the digital time span of the discriminated pulses. The individual channel threshold DAC (4-bit DAC) was fixed to its minimum value and the high voltage adjustment  DAC (8-bit DAC) was set following the procedure detailed in the subsection~\ref{subsec:Equalization Between Channels} (equalization). The common chip threshold adjustment DAC (10-bit DAC) is used to set different discrimination levels. The third stage is a wide-band amplifier with a gain value of ten, to allow the measurement of the analog signal of the SiPM. The fourth stage consists of a 4\,m plastic scintillation bar with a threaded 5\,m wavelength-shifting optical fiber, built identically to the scintillator strips that conform an AMIGA muon counter. At the end of the optical fiber there is an optical connector coupled to the SiPM. The extra meter of fiber is between the scintillator and the SiPM. 
The fifth stage is an ad-hoc muon telescope trigger~\cite{paper_scanner} which triggers the acquisition every time a particle passes through each position of the main scintillating bar where the telescope is placed. It consists of two identical parts, each of them made of a scintillator of dimensions 4\,cm x 4\,cm x 1\,cm with a SiPM coupled for light detection. Both pieces are aligned above and below the main scintillating bar. The separation distance between triggers is 5\,cm and the main scintillating bar is placed equidistantly between both parts. This configuration selects a particular solid angle of the muons that impinges the main scintillator.   
The trigger condition is overcome when the two SiPMs, up and down, record signals above a given threshold in a time-coincidence window of 60\,ns. There is a probability of $\sim$2\,\% of false coincidences of this muon telescope mainly due to three factors: the probability of two different particles impinging each scintillator simultaneously without passing, neither of them, through the main scintillator, misalignment of the telescope with respect to the main scintillator, and simultaneous dark noise at both devices. 
The sixth stage is the acquisition system. This stage is composed of a Tektronix DPO7104 oscilloscope. This oscilloscope is set up to store the discriminated signal of the CITIROC (second stage) and the amplified analog signal of the SiPM (third stage) every time the muon telescope produces a coincidence in a time window of 60\,ns (fifth stage).

\subsubsection{Selection of the Counting Strategy}
\label{subsec:counting_strategy}

As was described in~\cite{ICRCWundheiler2015}, the current counting system (PMT and electronics) has been designed to count muons by identifying a pattern in the digital trace of a fixed length of 3.2\,$\mu$s. The length of the digital trace is fixed by the cosmic ray shower duration. The discrimination level is set to a value lower than one SPE and then a pattern recognition technique is applied to discriminate particles from noise. This counting strategy takes into account the time structure of the signal over a threshold. 

The proposed counting strategy for the SiPMs is based on an amplitude criteria. The high noise of SiPMs implies setting a threshold of a small number of SPEs to discriminate particles from noise. The high PDE ($\sim$35\,\%) of SiPM at the emission wavelength of the WLS optical fiber mitigates the loss of detection efficiency.

\begin{figure}[h!t]
	\centering
	\includegraphics[width=0.65\textwidth]{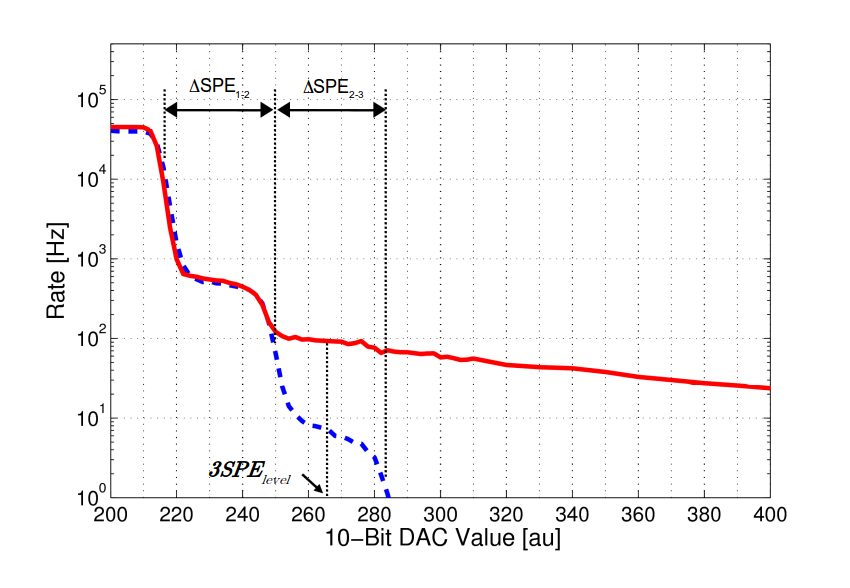}
	\caption{Measurement of the SiPM pulse rate with (red complete line) and without (blue dashed line) the scintillating bar over the 10-bit DAC value. Each plateau represents the transition for different amounts of SPEs. The value marked with the $3SPE_{level}$ is the selected value for the threshold of the discriminator. All the measurements were done at 25\,$^{\circ}$C and $\Delta V = 3.75\,V$.}
	\label{fig:dr_muon}
\end{figure}

The discrimination level is set to the lowest value that ensures a low rate of contamination (negligible accidental counting) without significantly reducing the counting efficiency. In figure~\ref{fig:dr_muon}, the rate of SiPM pulses as a function of the 10-bit DAC threshold is shown. For these measurements the muon telescope was not used and the rate is acquired directly from the CITIROC digital output with a pulse counter programmed in the FPGA. Two cases are plotted: the red complete line is the rate when the fiber is coupled to the SiPM, and the blue dashed line when it is not. When the fiber is not coupled to the SiPM, only the noise from the SiPM is measured. When the fiber is coupled to the SiPM, not only the dark rate and its correlated noise are measured, but also all the signals produced by charged particles impinging the scintillator. These particles will be considered as the environmental radiation. This environmental radiation is composed of cosmic ray particles (muons and others) and radiation of the surrounding materials (e.g. walls, floor, casing). In figure~\ref{fig:dr_muon}, for one and two SPE rate levels, the correlated and the uncorrelated noises, explained in section~\ref{subsec:SiPM Noise}, dominate. If the threshold level is set to any of these values, the accidental counting probability for the whole detector (64 channels) is over the desired limit level of 5\,\%. This 5\,\% was selected considering that the overcounting must remain below the error value of the number of muons reported in~\citep{paper_muon_lateral_distribution}, see Figures 5, 8, and 11. At three SPE rate level, the environmental radiation starts to dominate over the noise. For this level, the accidental counting probability (see equation~\ref{eq:accidental counting}) is mainly due to the environmental radiation and fulfills the requirement, therefore this is the selected level for the threshold. 

As it was mentioned in subsection~\ref{subsubsec:Single Photon Equivalent Peak Measurement}, the transition between plateaus in the SiPM rate as a function of the 10-bit DAC value represents the threshold of the comparator passing through the SPE peaks. Furthermore, as was mentioned in section~\ref{sec:The Silicon Photomultiplier}, the amplitude of the signal generated by multiple simultaneously fired cells is directly proportional to the number of them. Therefore, it is expected that: $\Delta SPE_{1-2} = \Delta SPE_{2-3} = SPE_{peak}$ (see figure~\ref{fig:dr_muon}). In the example shown, the obtained values were:  $\Delta SPE_{1-2} = (32.0 \pm 1.4);\ \Delta SPE_{2-3} = (36.0 \pm 1.4);\ SPE_{peak} = (34.0 \pm 1.4)$ (in arbitrary units).

The middle point of the transition from the second to the third SPE peaks ($SPE_{peak} \ast 2.5$) corresponds to the value that ensures the detection of signals with three or more SPEs ($3SPE_{level}$). Considering the offset ($DC_{level}$), this value can be calculated with the following equation:
\begin{equation}
\begin{split}
3SPE_{level}& = SPE_{peak} \ast 2.5 + DC_{level} \simeq ((216.0 \pm 1.0)-(182.0 \pm 1.0)) \ast 2.5 + (182.0 \pm 1.0)\\
& = (216.0 \pm 1.0)\ast 2.5 -(182.0 \pm 1.0)\ast 1.5 = 267.0 \pm 2.9
\end{split}
\label{eq:3SPEdiscriminator}
\end{equation}
As an example, in the case exemplified in figure~\ref{fig:dr_muon} the estimated $SPE_{peak}$ was $(34.0 \pm 1.4)$ and the $DC_{level}$ was $(182.0 \pm 1.0)$. The calculated $3SPE_{level}$ value is indicated in the figure, and it corresponds to the middle point of the third SPE plateau. The $3SPE_{level}$ must be estimated and set individually for each channel.

To estimate the accidental counting probability ($P_{accidental-counting}$) three factors are taken into account: the segmentation or number of strips ($n$), the acquired event time window ($T_{event}$) and the noise rate ($R_{noise}$, i.e. the environmental radiation and dark rate). As an example, for the AMIGA modules the segmentation is of 64 channels, the acquired event time window is 3.2\,$\mu$s and the noise rate in the underground laboratory is $\sim$100\,Hz (this value is the rate corresponding to $3SPE_{level}$ in figure~\ref{fig:dr_muon}). 
With those values, the accidental counting probability for a 10\,m$^2$ module was estimated to be:
\begin{equation}
P_{accidental-counting} = n \cdot T_{event} \cdot R_{noise} \simeq 64 \cdot 3.2\,\mu s \cdot 100\,Hz \equiv 2\,\%
\label{eq:accidental counting}
\end{equation}
which is below the desired level of 5\,\%.
The efficiency will be studied in detail in section~\ref{sec:Efficiency Measurements}.

\subsection{Proposed On-site Calibration}

The electronics design enables calibration to be performed at the observatory site. To ensure its long-term performance, both calibrations detailed in sections~\ref{subsec:CalibSiPM} and~\ref{subsec:CalibDetector} will be applied regularly and automatically. Each module will acquire the calibration data locally and then transmit it to a dedicated calibration server that will be running in the Central Data Acquisition System (CDAS) of the Pierre Auger Observatory. The calibration server will carry out the SiPM calibration as well as the detector calibration. This dedicated server will do the calculations to set three groups of parameters: the HV value, the 8-bit DAC to equalize the channels, and the 10-bit DAC value to set the discrimination level. All the calibration data and the parameters obtained will be stored for long-term stability studies. 

\section{Counting Efficiency Measurements}
\label{sec:Efficiency Measurements}

As mentioned in section~\ref{sec:Intro}, the module must count efficiently the number of impinging muons. To test its efficiency, the setup described in subsection~\ref{subsec:Detector Calibration Setup} was used. Several measurements at different fiber lengths were taken using the muon telescope. 
Every time there is a muon telescope trigger (event), the acquisition system (see sixth stage of section~\ref{subsec:Detector Calibration Setup}) stores the discriminated signal of the CITIROC (digital trace) and the amplified analog signal of the SiPM (analog trace). 

\begin{figure}[h!t]
	\centering
	\includegraphics[width=0.95\textwidth]{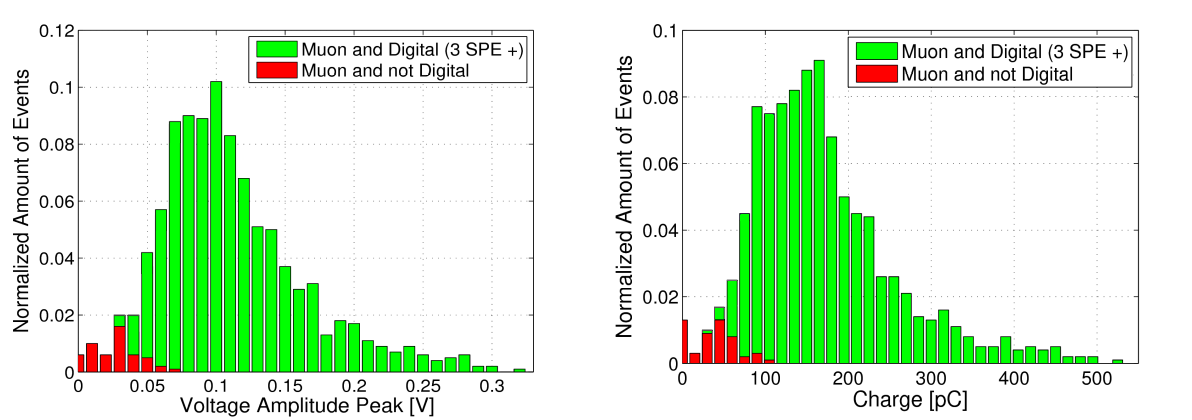}
	\caption{In the left (right) plot, the voltage amplitude peak (charge) histogram of 1000 triggered muon events is shown. The red colored area corresponds to the events that do not have a positive digital trace. All the measurements were done at 25\,$^{\circ}$C, $\Delta V = 3.75\,V$, and with the muon telescope placed at 430\,cm of fiber from the SiPM.}
	\label{fig:pico_carga}
\end{figure}

From the amplified analog traces, the voltage amplitude peak and charge (integral of the current) of each pulse were obtained by an offline analysis. The charge was numerically calculated integrating the voltage signal trace, in a time window of 200\,ns, divided by the oscilloscope input impedance value (50\,$\Omega$). A histogram illustrating the voltage amplitude peak and charge of the analog traces obtained at a fiber distance of 430\,cm with the S13081-050CS SiPM is shown in figure~\ref{fig:pico_carga}. Two colored areas can be distinguished in each histogram. The red area represents the events that do not have a corresponding positive digital trace of the CITIROC.  This means that no particle was detected by the counting system. The green area represents all the events that have a positive digital trace. 

\begin{figure}[h!t]
	\centering
	\includegraphics[width=0.95\textwidth]{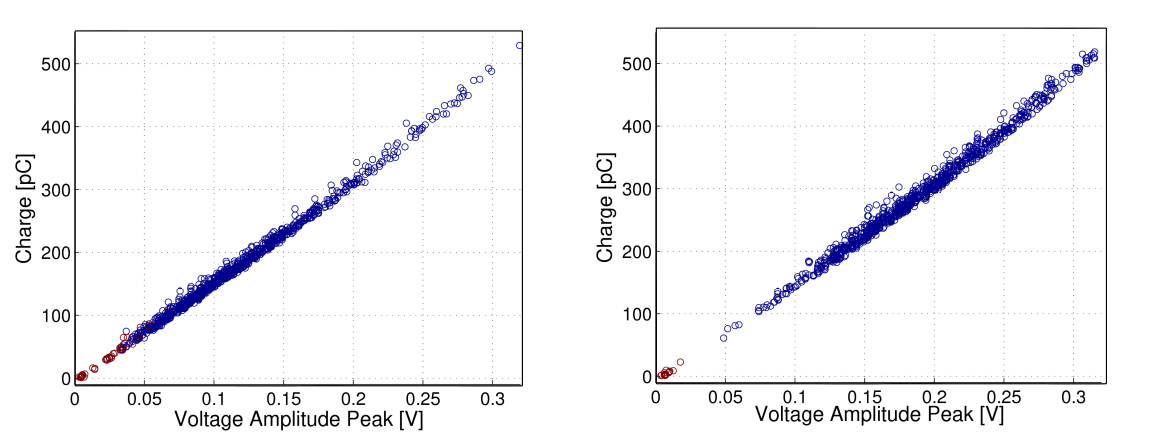}
	\caption{Both plots show the relationship between the voltage amplitude peak and the charge of 1000  triggered muon events. The red colored points correspond to events that do not have a digital output. In the left (right) plot the muon telescope was placed at 430\,cm (130\,cm) of fiber. Isolated red points in the lower left corner in the right figure evidence the false coincidences of the muon telescope trigger. All the measurements were done at 25\,$^{\circ}$C and $\Delta V = 3.75\,V$.}  
	\label{fig:peak_vs_charge}
\end{figure}

Both plots in figure~\ref{fig:peak_vs_charge} show a correlation between the voltage amplitude peak and charge of SiPM signals. In these plots there is a discrimination between the traces with (blue) or without (red) digital output. In the left plot, the results with the muon telescope placed at 430\,cm of fiber are shown, and in the right one, the muon telescope was moved to 130\,cm. As expected, the data sets have higher mean voltage amplitude and charge values due to a decrease in the light attenuation of the fiber. 

\begin{figure}[h!t]
	\centering
	\includegraphics[width=0.95\textwidth]{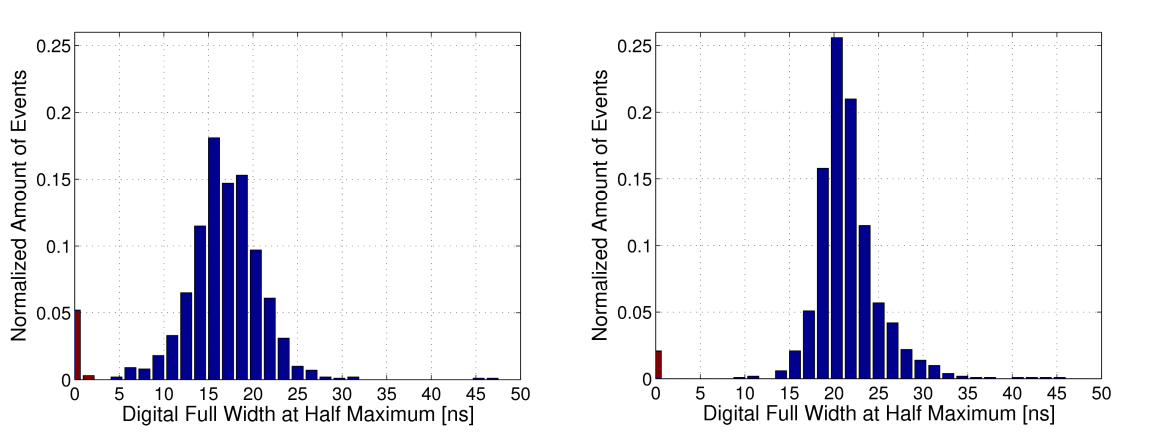}
	\caption{Both plots show the width histogram of 1000 digital output pulses of the CITIROC. The red colored bar corresponds to the traces that do not have a digital output pulse. In the left (right) plot the muon telescope was placed at 430\,cm (130\,cm) of fiber. All the measurements were done at 25\,$^{\circ}$C and $\Delta V = 3.75\,V$.}
	\label{fig:digital_width}
\end{figure}

In figure~\ref{fig:digital_width}, the full width at half maximum (FWHM) of the digital output pulse of the CITIROC, acquired with the oscilloscope at two fiber distances measured is shown. As it was pointed out in the requirements, $99\,\%$ of the digital widths are lower than 35\,ns. This requirement was achieved by the fast shaper included in the CITIROC. There was no evidence of afterpulses in the observed digital output. 

\begin{figure}[h!t]
	\centering
	\includegraphics[width=0.65\textwidth]{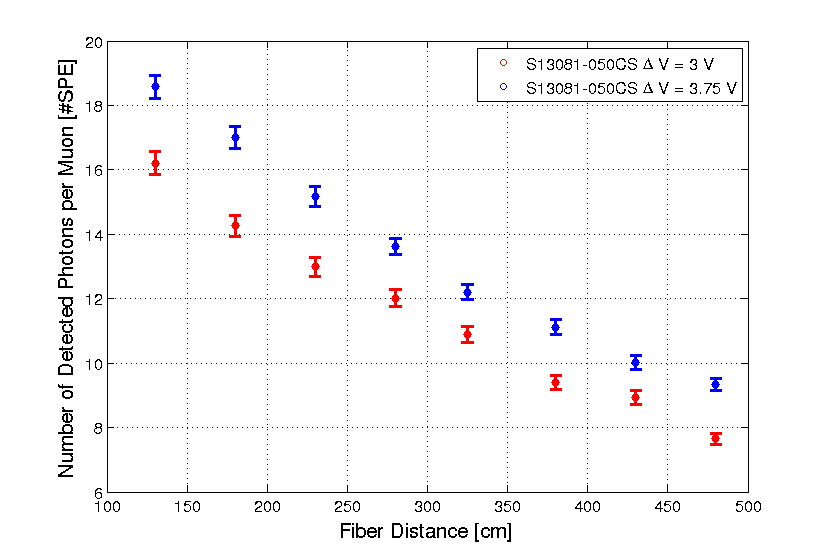}
	\caption{Number of detected photons per muon impinging the detector as a function of the fiber distance for two different overvoltages: $\Delta V = 3\,V$ and $\Delta V = 3.75\,V$. The detector light yield is not uniform due to the fiber attenuation. Error bars are only due to statistic errors of the mean. All the measurements were done at 25\,$^{\circ}$C.}
	\label{fig:attenuation}
\end{figure}

In figure~\ref{fig:attenuation}, the mean value of the number of detected photons per muon impinging the detector as a function of the fiber distance is shown. This mean value was obtained as the mean value of the distribution of the ratio between the charge of measured traces and the mean charge of the SPE. Error bars are only due to statistic errors of the mean. This light yield curve was obtained with the setup shown in Figure~\ref{fig:Detector_Calibration_Setup}. The muon telescope trigger selects mainly vertical muons and the acceptance angle considering the muon telescope geometry is $\rm \sim 35\,^{\circ}$. There are two main factors that constrain the performance of the detector since its light yield is not uniform due to the fiber attenuation. In the farther distances, the efficiency strongly depends on the threshold selection, since the attenuation of the optical fiber significantly decreases the number of photons that arrive to the SiPM. At the closest distances, the number of detected photons is higher and the digital width is consequently increased. The selected pre-amplifier gain, i.e. high-gain pre-amplifier set to the lowest gain, combined with the fast shaper ensures an adequate width of the digital output. 

\subsection{Efficiency Results and Possible Improvements}

The counting efficiency is defined as the ratio between the amount of positive digital output traces of the CITIROC, and the number of triggers of the muon telescope that ensures a particle passing through the scintillating bar at a certain distance.

\begin{figure}[h!t]
	\centering
	\includegraphics[width=0.65\textwidth]{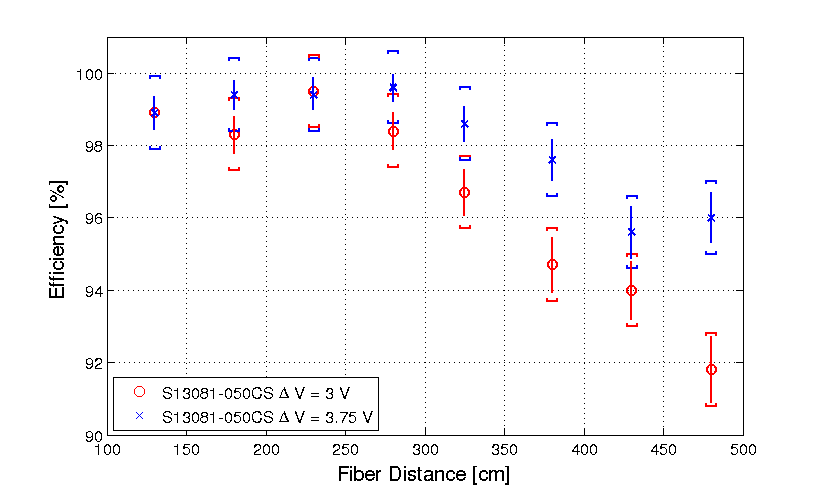}
	\caption{Efficiency measurements for two different $\Delta V$ are compared. A larger $\Delta V$ produces an increase in the efficiency. The estimated integrated efficiency is: 97\,\% for $\Delta V = 3\,V$ (red) and 98\,\% for $\Delta V = 3.75\,V$ (blue). Vertical lines represent statistic errors assuming a binomial distribution and brackets represent the systematics errors caused by false triggers.} 
	\label{fig:efficiency}
\end{figure}

Figure~\ref{fig:efficiency} summarizes the results of the efficiency for eight different distances and for two different overvoltages. Two kind of errors are included in the plot. Vertical lines represent statistic errors assuming a binomial distribution and brackets represent the systematic errors caused by false triggers. It should be pointed out that the efficiency study was performed with the setup shown in Figure~\ref{fig:Detector_Calibration_Setup}. The muon telescope trigger selects mainly vertical muons, the acceptance angle considering the muon telescope geometry is $\rm \sim 35\,^{\circ}$.

A higher efficiency is achieved when the $\Delta V$ is increased, since the PDE of the SiPM rises with the increment of the $\Delta V$. With the increment of the $\Delta V$, the noise is also increased. For the higher $\Delta V$, the accidental counting probability is $\sim$2\,\% (see section~\ref{subsec:counting_strategy}).

Although the efficiency has a dependence with the distance, one important parameter that should be calculated is the integrated efficiency over the whole scintillating bar. For the two cases exemplified in figure~\ref{fig:efficiency}, the estimated integrated efficiency is: 97\,\% for $\Delta V = 3\,V$ and 98\,\% for $\Delta V = 3.75\,V$. 

\section{Conclusions}

A new readout for the AMIGA muon counters was proposed. The selected SiPM was the S13081-050CS due to its low crosstalk and afterpulsing. The CITIROC ASIC was selected as the electronics front-end. 
Its fast shaper enables a digital output width of the discriminator similar to the characteristic time width of the light pulses produced by the impinging particles in the detector. The Hamamatsu C11204-01 power supply was chosen for the biasing of the SiPMs.  

The proposed calibration method consists of two steps. Firstly, the SiPM calibration allows the individual characterization of each SiPM of the module and the equalization between channels. Secondly, the detector calibration determines the discrimination level and the counting strategy. Both calibrations combined guarantee the performance of the detector by an adequate overvoltage and threshold level selection. Both methods were designed to be performed on the Observatory site. This allows studying the long-term performance of the detector, improving its stability for long periods.

Laboratory efficiency studies show promising results. The high integrated efficiency obtained (98\,\% for the higher tested overvoltage) combined with a low probability of accidental counting ($\sim$2\,\%) evidences an adequate performance of the proposed counting system.


\section*{Acknowledgments}

\begin{sloppypar}
The successful installation, commissioning, and operation of the Pierre Auger Observatory would not have been possible without the strong commitment and effort from the technical and administrative staff in Malarg\"ue. We are very grateful to the following agencies and organizations for financial support:
\end{sloppypar}

\begin{sloppypar}
Comisi\'on Nacional de Energ\'\i{}a At\'omica, Agencia Nacional de Promoci\'on Cient\'\i{}fica y Tecnol\'ogica (ANPCyT), Consejo Nacional de Investigaciones Cient\'\i{}ficas y T\'ecnicas (CONICET), Gobierno de la Provincia de Mendoza, Municipalidad de Malarg\"ue, NDM Holdings and Valle Las Le\~nas, in gratitude for their continuing cooperation over land access, Argentina; the Australian Research Council; Conselho Nacional de Desenvolvimento Cient\'\i{}fico e Tecnol\'ogico (CNPq), Financiadora de Estudos e Projetos (FINEP), Funda\c{c}\~ao de Amparo \`a Pesquisa do Estado de Rio de Janeiro (FAPERJ), S\~ao Paulo Research Foundation (FAPESP) Grants No.\ 2010/07359-6 and No.\ 1999/05404-3, Minist\'erio de Ci\^encia e Tecnologia (MCT), Brazil; Grant No.\ MSMT CR LG15014, LO1305 and LM2015038 and the Czech Science Foundation Grant No.\ 14-17501S, Czech Republic; Centre de Calcul IN2P3/CNRS, Centre National de la Recherche Scientifique (CNRS), Conseil R\'egional Ile-de-France, D\'epartement Physique Nucl\'eaire et Corpusculaire (PNC-IN2P3/CNRS), D\'epartement Sciences de l'Univers (SDU-INSU/CNRS), Institut Lagrange de Paris (ILP) Grant No.\ LABEX ANR-10-LABX-63, within the Investissements d'Avenir Programme Grant No.\ ANR-11-IDEX-0004-02, France; Bundesministerium f\"ur Bildung und Forschung (BMBF), Deutsche Forschungsgemeinschaft (DFG), Finanzministerium Baden-W\"urttemberg, Helmholtz Alliance for Astroparticle Physics (HAP), Helmholtz-Gemeinschaft Deutscher Forschungszentren (HGF), Ministerium f\"ur Wissenschaft und Forschung, Nordrhein Westfalen, Ministerium f\"ur Wissenschaft, Forschung und Kunst, Baden-W\"urttemberg, Germany; Istituto Nazionale di Fisica Nucleare (INFN),Istituto Nazionale di Astrofisica (INAF), Ministero dell'Istruzione, dell'Universit\'a e della Ricerca (MIUR), Gran Sasso Center for Astroparticle Physics (CFA), CETEMPS Center of Excellence, Ministero degli Affari Esteri (MAE), Italy; Consejo Nacional de Ciencia y Tecnolog\'\i{}a (CONACYT) No.\ 167733, Mexico; Universidad Nacional Aut\'onoma de M\'exico (UNAM), PAPIIT DGAPA-UNAM, Mexico; Ministerie van Onderwijs, Cultuur en Wetenschap, Nederlandse Organisatie voor Wetenschappelijk Onderzoek (NWO), Stichting voor Fundamenteel Onderzoek der Materie (FOM), Netherlands; National Centre for Research and Development, Grants No.\ ERA-NET-ASPERA/01/11 and No.\ ERA-NET-ASPERA/02/11, National Science Centre, Grants No.\ 2013/08/M/ST9/00322, No.\ 2013/08/M/ST9/00728 and No.\ HARMONIA 5 -- 2013/10/M/ST9/00062, Poland; Portuguese national funds and FEDER funds within Programa Operacional Factores de Competitividade through Funda\c{c}\~ao para a Ci\^encia e a Tecnologia (COMPETE), Portugal; Romanian Authority for Scientific Research ANCS, CNDI-UEFISCDI partnership projects Grants No.\ 20/2012 and No.194/2012 and PN 16 42 01 02; Slovenian Research Agency, Slovenia; Comunidad de Madrid, Fondo Europeo de Desarrollo Regional (FEDER) funds, Ministerio de Econom\'\i{}a y Competitividad, Xunta de Galicia, European Community 7th Framework Program, Grant No.\ FP7-PEOPLE-2012-IEF-328826, Spain; Science and Technology Facilities Council, United Kingdom; Department of Energy, Contracts No.\ DE-AC02-07CH11359, No.\ DE-FR02-04ER41300, No.\ DE-FG02-99ER41107 and No.\ DE-SC0011689, National Science Foundation, Grant No.\ 0450696, The Grainger Foundation, USA; NAFOSTED, Vietnam; Marie Curie-IRSES/EPLANET, European Particle Physics Latin American Network, European Union 7th Framework Program, Grant No.\ PIRSES-2009-GA-246806; and UNESCO.
\end{sloppypar}

\newpage
\section*{The Pierre Auger Collaboration}
\addcontentsline{toc}{section}{The Pierre Auger Collaboration}
{\small

A.~Aab$^{37}$,
P.~Abreu$^{69}$,
M.~Aglietta$^{47,46}$,
E.J.~Ahn$^{84}$,
I.~Al Samarai$^{29}$,
I.F.M.~Albuquerque$^{16}$,
I.~Allekotte$^{1}$,
P.~Allison$^{89}$,
A.~Almela$^{8,11}$,
J.~Alvarez Castillo$^{61}$
% ...


\begin{description}[labelsep=0.2em,align=right,labelwidth=0.7em,labelindent=0em,leftmargin=2em,noitemsep]
\item[$^{1}$] Centro At\'omico Bariloche and Instituto Balseiro (CNEA-UNCuyo-CONICET), Argentina
\item[$^{2}$] Centro de Investigaciones en L\'aseres y Aplicaciones, CITEDEF and CONICET, Argentina
\item[$^{3}$] Departamento de F\'\i{}sica and Departamento de Ciencias de la Atm\'osfera y los Oc\'eanos, FCEyN, Universidad de Buenos Aires, Argentina
\item[$^{4}$] IFLP, Universidad Nacional de La Plata and CONICET, Argentina
\item[$^{5}$] Instituto de Astronom\'\i{}a y F\'\i{}sica del Espacio (IAFE, CONICET-UBA), Argentina
\item[$^{6}$] Instituto de F\'\i{}sica de Rosario (IFIR) -- CONICET/U.N.R.\ and Facultad de Ciencias Bioqu\'\i{}micas y Farmac\'euticas U.N.R., Argentina
\item[$^{7}$] Instituto de Tecnolog\'\i{}as en Detecci\'on y Astropart\'\i{}culas (CNEA, CONICET, UNSAM) and Universidad Tecnol\'ogica Nacional -- Facultad Regional Mendoza (CONICET/CNEA), Argentina
\item[$^{8}$] Instituto de Tecnolog\'\i{}as en Detecci\'on y Astropart\'\i{}culas (CNEA, CONICET, UNSAM), Centro At\'omico Constituyentes, Comisi\'on Nacional de Energ\'\i{}a At\'omica, Argentina
\item[$^{9}$] Observatorio Pierre Auger, Argentina
\item[$^{10}$] Observatorio Pierre Auger and Comisi\'on Nacional de Energ\'\i{}a At\'omica, Argentina
\item[$^{11}$] Universidad Tecnol\'ogica Nacional -- Facultad Regional Buenos Aires, Argentina
\item[$^{12}$] University of Adelaide, Australia
\item[$^{13}$] Centro Brasileiro de Pesquisas Fisicas (CBPF), Brazil
\item[$^{14}$] Universidade de S\~ao Paulo, Escola de Engenharia de Lorena, Brazil
\item[$^{15}$] Universidade de S\~ao Paulo, Inst.\ de F\'\i{}sica de S\~ao Carlos, S\~ao Carlos, Brazil
\item[$^{16}$] Universidade de S\~ao Paulo, Inst.\ de F\'\i{}sica, S\~ao Paulo, Brazil
\item[$^{17}$] Universidade Estadual de Campinas (UNICAMP), Brazil
\item[$^{18}$] Universidade Estadual de Feira de Santana (UEFS), Brazil
\item[$^{19}$] Universidade Federal de Pelotas, Brazil
\item[$^{20}$] Universidade Federal do ABC (UFABC), Brazil
\item[$^{21}$] Universidade Federal do Paran\'a, Setor Palotina, Brazil
\item[$^{22}$] Universidade Federal do Rio de Janeiro (UFRJ), Instituto de F\'\i{}sica, Brazil
\item[$^{23}$] Universidade Federal Fluminense, Brazil
\item[$^{24}$] Universidad Industrial de Santander, Colombia
\item[$^{25}$] Institute of Physics (FZU) of the Academy of Sciences of the Czech Republic, Czech Republic
\item[$^{26}$] Palacky University, RCPTM, Czech Republic
\item[$^{27}$] University Prague, Institute of Particle and Nuclear Physics, Czech Republic
\item[$^{28}$] Institut de Physique Nucl\'eaire d'Orsay (IPNO), Universit\'e Paris 11, CNRS-IN2P3, France
\item[$^{29}$] Laboratoire de Physique Nucl\'eaire et de Hautes Energies (LPNHE), Universit\'es Paris 6 et Paris 7, CNRS-IN2P3, France
\item[$^{30}$] Laboratoire de Physique Subatomique et de Cosmologie (LPSC), Universit\'e Grenoble-Alpes, CNRS/IN2P3, France
\item[$^{31}$] Bergische Universit\"at Wuppertal, Department of Physics, Germany
\item[$^{32}$] Karlsruhe Institute of Technology, Institut f\"ur Experimentelle Kernphysik (IEKP), Germany
\item[$^{33}$] Karlsruhe Institute of Technology, Institut f\"ur Kernphysik (IKP), Germany
\item[$^{34}$] Karlsruhe Institute of Technology, Institut f\"ur Prozessdatenverarbeitung und Elektronik (IPE), Germany
\item[$^{35}$] RWTH Aachen University, III.\ Physikalisches Institut A, Germany
\item[$^{36}$] Universit\"at Hamburg, II.\ Institut f\"ur Theoretische Physik, Germany
\item[$^{37}$] Universit\"at Siegen, Fachbereich 7 Physik -- Experimentelle Teilchenphysik, Germany
\item[$^{38}$] INAF -- Istituto di Astrofisica Spaziale e Fisica Cosmica di Palermo, Italy
\item[$^{39}$] INFN Laboratori del Gran Sasso, Italy
\item[$^{40}$] INFN, Sezione di Catania, Italy
\item[$^{41}$] INFN, Sezione di L'Aquila, Italy
\item[$^{42}$] INFN, Sezione di Lecce, Italy
\item[$^{43}$] INFN, Sezione di Milano, Italy
\item[$^{44}$] INFN, Sezione di Napoli, Italy
\item[$^{45}$] INFN, Sezione di Roma ``Tor Vergata``, Italy
\item[$^{46}$] INFN, Sezione di Torino, Italy
\item[$^{47}$] Osservatorio Astrofisico di Torino (INAF), Torino, Italy
\item[$^{48}$] Universit\`a del Salento, Dipartimento di Ingegneria, Italy
\item[$^{49}$] Universit\`a del Salento, Dipartimento di Matematica e Fisica ``E.\ De Giorgi'', Italy
\item[$^{50}$] Universit\`a dell'Aquila, Dipartimento di Chimica e Fisica, Italy
\item[$^{51}$] Universit\`a di Catania, Dipartimento di Fisica e Astronomia, Italy
\item[$^{52}$] Universit\`a di Milano, Dipartimento di Fisica, Italy
\item[$^{53}$] Universit\`a di Napoli ``Federico II``, Dipartimento di Fisica ``Ettore Pancini``, Italy
\item[$^{54}$] Universit\`a di Roma ``Tor Vergata'', Dipartimento di Fisica, Italy
\item[$^{55}$] Universit\`a Torino, Dipartimento di Fisica, Italy
\item[$^{56}$] Benem\'erita Universidad Aut\'onoma de Puebla (BUAP), M\'exico
\item[$^{57}$] Centro de Investigaci\'on y de Estudios Avanzados del IPN (CINVESTAV), M\'exico
\item[$^{58}$] Unidad Profesional Interdisciplinaria en Ingenier\'\i{}a y Tecnolog\'\i{}as Avanzadas del Instituto Polit\'ecnico Nacional (UPIITA-IPN), M\'exico
\item[$^{59}$] Universidad Aut\'onoma de Chiapas, M\'exico
\item[$^{60}$] Universidad Michoacana de San Nicol\'as de Hidalgo, M\'exico
\item[$^{61}$] Universidad Nacional Aut\'onoma de M\'exico, M\'exico
\item[$^{62}$] Institute for Mathematics, Astrophysics and Particle Physics (IMAPP), Radboud Universiteit, Nijmegen, Netherlands
\item[$^{63}$] KVI -- Center for Advanced Radiation Technology, University of Groningen, Netherlands
\item[$^{64}$] Nationaal Instituut voor Kernfysica en Hoge Energie Fysica (NIKHEF), Netherlands
\item[$^{65}$] Stichting Astronomisch Onderzoek in Nederland (ASTRON), Dwingeloo, Netherlands
\item[$^{66}$] Institute of Nuclear Physics PAN, Poland
\item[$^{67}$] University of \L{}\'od\'z, Faculty of Astrophysics, Poland
\item[$^{68}$] University of \L{}\'od\'z, Faculty of High-Energy Astrophysics, Poland
\item[$^{69}$] Laborat\'orio de Instrumenta\c{c}\~ao e F\'\i{}sica Experimental de Part\'\i{}culas -- LIP and Instituto Superior T\'ecnico -- IST, Universidade de Lisboa -- UL, Portugal
\item[$^{70}$] ``Horia Hulubei'' National Institute for Physics and Nuclear Engineering, Romania
\item[$^{71}$] Institute of Space Science, Romania
\item[$^{72}$] University of Bucharest, Physics Department, Romania
\item[$^{73}$] University Politehnica of Bucharest, Romania
\item[$^{74}$] Experimental Particle Physics Department, J.\ Stefan Institute, Slovenia
\item[$^{75}$] Laboratory for Astroparticle Physics, University of Nova Gorica, Slovenia
\item[$^{76}$] Universidad Complutense de Madrid, Spain
\item[$^{77}$] Universidad de Alcal\'a de Henares, Spain
\item[$^{78}$] Universidad de Granada and C.A.F.P.E., Spain
\item[$^{79}$] Universidad de Santiago de Compostela, Spain
\item[$^{80}$] Case Western Reserve University, USA
\item[$^{81}$] Colorado School of Mines, USA
\item[$^{82}$] Colorado State University, USA
\item[$^{83}$] Department of Physics and Astronomy, Lehman College, City University of New York, USA
\item[$^{84}$] Fermi National Accelerator Laboratory, USA
\item[$^{85}$] Louisiana State University, USA
\item[$^{86}$] Michigan Technological University, USA
\item[$^{87}$] New York University, USA
\item[$^{88}$] Northeastern University, USA
\item[$^{89}$] Ohio State University, USA
\item[$^{90}$] Pennsylvania State University, USA
\item[$^{91}$] University of Chicago, USA
\item[$^{92}$] University of Hawaii, USA
\item[$^{93}$] University of Nebraska, USA
\item[$^{94}$] University of New Mexico, USA
\item[] -----
\item[$^{a}$] School of Physics and Astronomy, University of Leeds, Leeds, United Kingdom
\item[$^{b}$] Max-Planck-Institut f\"ur Radioastronomie, Bonn, Germany
\item[$^{c}$] also at Vrije Universiteit Brussels, Brussels, Belgium
\item[$^{d}$] now at Deutsches Elektronen-Synchrotron (DESY), Zeuthen, Germany
\item[$^{e}$] SUBATECH, \'Ecole des Mines de Nantes, CNRS-IN2P3, Universit\'e de Nantes
\end{description}

}
\end{document}